\documentclass[prb,twocolumn,showpacs]{revtex4}
\usepackage[dvips]{graphicx}
\usepackage{epsfig}
\usepackage{amsmath,amsfonts,amssymb,bm}
\usepackage[caption=false]{subfig}
\setcitestyle{numbers,square}
\begin{document}
\title{Localization in a random $x-y$ model with the long-range interaction: Intermediate case between single particle and many-body problems}
\author{Alexander L. Burin}
\affiliation{Department of Chemistry, Tulane University, New
Orleans, LA 70118, USA}
\date{\today}
\begin{abstract}
Many-body localization in an $XY$ model with a long-range interaction  is investigated. We show that in the regime of a high strength of  disordering compared to the interaction an off-resonant flip-flop spin-spin interaction (hopping) generates the effective Ising interactions of spins in the third order of perturbation theory in a hopping. The combination of hopping and induced Ising interactions  for the power law distance dependent hopping $V(R) \propto R^{-\alpha}$  always leads to the localization breakdown in a thermodynamic limit of an infinite system at $\alpha < 3d/2$   where $d$ is a system dimension.  The delocalization takes place due to  the induced Ising interactions $U(R) \propto R^{-2\alpha}$ of ``extended" resonant pairs. This prediction is consistent with the numerical finite size scaling in one-dimensional systems. Many-body localization in $XY$ model is more stable with respect to the long-range interaction compared to  a many-body problem with similar  Ising and Heisenberg interactions requiring $\alpha \geq 2d$ which makes the practical implementations of this model more attractive for quantum information applications. The full summary of dimension constraints and localization threshold size dependencies for many-body localization in the case of combined Ising and hopping interactions is obtained using this and previous work and it is the subject for the future experimental verification using cold atomic systems. 
\end{abstract}

\maketitle


\section{Introduction}


Many-body localization-delocalization transition serves as a natural generalization of a single particle Anderson localization concept \cite{PW} to interacting quantum systems at finite temperature \cite{FleishmanAnderson1980,Kamenev} and as a quantum mechanical extension of a classical transition between deterministic and chaotic behaviors \cite{Dorf,Doron}. Localized systems show non-ergodic behavior where each small subsystem of it remembers its initial state during an infinite time. 
In the delocalized state the system serves as a thermal bath for each part of it (e. g. spin or particle) leading them to their thermal equilibrium. Many-body localization has been considered in a variety of physical systems \cite{KaganMaksimov,abKontor,Volynes,abmbl1,BerkovShkl,Basko06} and it attracts growing attention because of its significance in quantum informatics \cite{Polkovnikov,Imamo,BurrellOsborne}.  Indeed, in the regime of localization the system remembers its initial state infinitely long, while in chaotic state this information is quickly erased by irreversible dynamics.  Recently it has been suggested to investigate many-body localization using cold atomic systems \cite{Polkovnikov,Lukin1,Lukin2,Kondov}, which can emulate various models for localization-delocalization transitions. Particularly  the model considered in the present paper can be realized in diatomic alkali
systems \cite{Lukin2}.


Many-body interaction often leads to the single-particle localization breakdown because it opens new channels for energy or particle transport \cite{FleishmanAnderson1980,AronovReview,Yu,BurinMaksimovPolishchuk,MirlinMBDel,Basko06,abmbl1,Lukin2,abmbl2} (see however Refs. \cite{KaganMaksimov,abKontor,Markus} where the localizing effect of quasi-static interaction has been exploited). 
Particularly, the long-range interaction decreasing with the distance according to the power law  due to dipolar, magnetic or elastic forces dramatically  enlarges the number of delocalization pathways because of long-distance connections.

Indeed, even at zero temperature excitations are delocalized at arbitrarily strong disorder in an infinite system if the interaction  decreases  with the distance slower than  $R^{-d}$ (see Refs. \cite{PW,Levitov2}). At a finite temperature and in the presence of a long-range interparticle (Ising) interaction $U_{ij}S_{i}^{z}S_{j}^{z}$ (or $U_{ij}n_{i}n_{j}$ for quasiparticles) decreasing with the distance as  $U_{ij} \propto R_{ij}^{-\beta}$ the inevitable delocalization is expected at $\beta < 2d$ (see Refs. \cite{abmbl1,Lukin2,abmbl2} and the analysis of delocalization in Sec. \ref{sec:maxsize}). This strong dimension constraint results from the resonant energy exchange  between the flip-flop transitions of spin pairs caused by Ising many-body interactions 
(see Fig. \ref{fig:ResPairs}). However there is no such interaction in the $XY$ model containing only hopping terms $V_{ij}S_{i}^{+}S_{j}^{-}$ (below we refer to this term as a {\it hopping interaction} emphasizing its both hopping and interaction nature). Therefore it is unclear how the long-range hopping would affect the localization there. 

To address this fundamental question and fill the existing gap in the dimensional constraints obtained only in the case of dominating Ising interaction \cite{abmbl1,Lukin2,abmbl2}  we investigate the effect of the long-range hopping interaction on the many-body localization in a random strongly disordered $XY$ model for interacting spins $1/2$.  We show that the effective Ising interaction between spins still exists and it is generated in the third order of perturbation theory in the hopping interaction $V_{ij}$. This interaction decreases with the distance as $R^{-2\alpha}$ and it leads to the localization breakdown for $\alpha < 3d/2$ in agreement with the ``extended pair" criterion of Ref. \cite{Lukin2}. 

An $XY$ model is relevant for a variety of phenomena including energy transport in  Josephson junction arrays \cite{IoffeFeigelman}, exciton transport in quantum gases \cite{Lukin1,Lukin2}  and many other problems of interest.  This model is also relevant for quantum informatics including quantum information processing using quantum dot spins \cite{Imamo} and quantum information transport \cite{BurrellOsborne}.  Therefore it is important to characterize the  dynamics in this model particularly understanding the specific of many-body localization there.

The paper is organized as following. The model is formulated in Sec. \ref{sec:Mod}. The effective Ising interaction is derived in Sec. \ref{sec:Ising} and  the many-body localization problem in the presence of natural and induced interactions is considered in Sec. \ref{sec:MBLXY}. The numerical verification of the results using the finite size scaling method is reported in Sec. \ref{sec:num}. The conclusions are formulated in Sec. \ref{sec:Conclusion}.



\section{Model}
\label{sec:Mod}

In this paper we investigate a many-body localization in an $XY$ model of $N$ interacting spins $1/2$ 
\begin{eqnarray}
\widehat{H}= \sum_{i}\phi_{i}S_{i}^{z}+\sum_{i \neq j} V_{ij}S_{i}^{+}S_{j}^{-},
\label{eq:H}
\end{eqnarray} 
occupying $d$-dimensional hypercube with the density $n$. Fields $\phi_{i}$ are not correlated in different sites $i$ and they are uniformly distributed within the domain $(-W/2, W/2)$. The breakdown of many-body localization is investigated for the hopping interaction decreasing with the distance according to the power law 
\begin{eqnarray}
V_{ij}\sim \frac{\tilde{V}}{(n^{\frac{1}{d}}R_{ij})^{\alpha}} 
\label{eq:IntHop}
\end{eqnarray}
where $\tilde{V}$ estimates the hopping  interaction at the average distance. 


We assume that the  disorder is strong $\tilde{V}\ll W$ so the only long-range interaction can lead to  delocalization. The consideration is restricted to the case $\alpha \geq d$ so the single-particle delocalization can be neglected (see discussion in Sec. \ref{sec:alph=d} for the threshold case $\alpha=d$). Accordingly there is no mobility edge  for the single particle states which are all strongly localized (cf. Ref.  \cite{Lazarides}). We consider the infinite temperature limit similarly to the previous work \cite{Reichmann,OgHuse1} because it is more convenient for analytical and numerical considerations while its generalization to the finite non-zero temperature is straightforward \cite{Comment1}.

In contrast with the previously considered models of interacting spins \cite{abmbl1,Lukin2,abmbl2}  with mixed interactions 
\begin{eqnarray}
\widehat{H}_{mixed}= \sum_{i}\phi_{i}S_{i}^{z}+\sum_{i \neq j} V_{ij}S_{i}^{+}S_{j}^{-}+\frac{1}{2}\sum_{i \neq j} U_{ij}S_{i}^{z}S_{j}^{z},
\label{eq:Hpr}
\end{eqnarray} 
Eq. (\ref{eq:H}) lacks the Ising  spin-spin interaction $U_{ij}$. In addition to the pure $XY$ model we will also discuss the general model Eq. (\ref{eq:Hpr}) in Sec. \ref{sec:genalphlessbeta} with the special attention to the regime $\alpha<\beta$ not considered yet. To make our consideration as general as possible  we introduce the second independent interaction parameter $\tilde{U}=Un^{\frac{\beta}{d}}$ (cf. Eq. (\ref{eq:IntHop})) representing the Ising interaction at the average distance. The large strength of disorder $\tilde{U}\ll W$ is also assumed for this interaction.

The  Fermion based version of the Hamiltonian Eq. (\ref{eq:Hpr}) can be generated replacing spin operators $S^{+}$, $S^{-}$, $S^{z}$ with Fermi-operators $a^{\dagger}$, $a$, $n=a^{\dagger}a$ (see Ref. \cite{Levitov4}) as
\begin{eqnarray}
\widehat{H}= \sum_{i}\phi_{i}n_{i}+\sum_{i \neq j} V_{ij}a_{i}^{\dagger}a_{j} + \frac{1}{2}\sum_{i \neq j} U_{ij}n_{i}n_{j}. 
\label{eq:HF}
\end{eqnarray} 
The latter model lacks many-body behavior in the absence of interparticle interaction  ($U_{ij}=0$) and it can be described in terms of independent Fermions which are localized in the case of a large strength of disorder for $\alpha<d$. The $XY$ model Eq. (\ref{eq:H}) behaves identically for the one dimensional problem with the nearest neighbor interaction \cite{JW28}, but shows different behavior if there exists triples of spins $i, j, k$ coupled by non-zero interactions $V_{ij}$, $V_{jk}$ and $V_{ik}$ as shown below in Sec.  \ref{sec:Ising}.

\section{Induced Ising interaction in XY model}
\label{sec:Ising}

\subsection{Motivation and zeroth order approximation}

Here we show that the effective Ising interaction exists in the $XY$ model with the long-range interaction. It is generated in the third order of perturbation theory in hopping interactions under conditions of a large strength of disorder (see Fig. \ref{fig:ResPairs} and derivation below). This interaction is relatively weak; yet it can lead to the delocalization. 

The  main reason for that is the dramatic significance of the Ising interaction for the many-body energy transport \cite{abmbl1,Lukin2,abmbl2}. Indeed the hopping interaction $V_{ij}S_{i}^{+}S_{j}^{-}$ bounds only sites with close random fields $\phi_{i}\approx \phi_{i'}$ while the Ising interaction is capable to induce the energy transport between  pairs of such sites $i, i'$ and $j, j'$ bound by the resonant condition  $\phi_{i} - \phi_{i'} \approx  \phi_{j} - \phi_{j'}$ while energies $\phi_{i}$ and $\phi_{j}$ can be very far from each other (see Fig. \ref{fig:ResPairs}). 
This transport channel can lead to the breakdown of many-body localization (see Ref. \cite{abmbl1} and Sec. IV). 
In addition it is fundamentally interesting to investigate the appearance of the difference between the random $XY$ model Eq. (\ref{eq:H}) and its Fermion counterpart Eq. (\ref{eq:HF}). The third order effect that we study is the first non-vanishing contribution to this difference.

To find the induced Ising interaction the perturbation theory is developed for the system  Eq. (\ref{eq:H}). One can separate the Hamiltonian into two parts including the random field term dominating in the case of a large strength of disorder
\begin{eqnarray} 
\widehat{H}_{0}= \sum_{i}\phi_{i}S_{i}^{z}
\label{eq:H0}
\end{eqnarray}
and considered as a zeroth order approximation and the hopping interaction term 
\begin{eqnarray} 
\widehat{V}= \sum_{i \neq j} V_{ij}S_{i}^{+}S_{j}^{-}
\label{eq:V}
\end{eqnarray}
 treated as a perturbation. This perturbation can be separated itself into resonant and non-resonant parts depending on whether the change of random field energy $\phi_{i}-\phi_{j}$ due to flip-flop transition is large or small compared to the hopping amplitudes $V_{ij}$.  Under conditions of strong disordered system $\tilde{V}\ll W$ one can introduce an intermediate crossover energy $\phi_{*}$, such that $\tilde{V} \ll \phi_{*} \ll W$ and treat the interaction part 
 \begin{eqnarray} 
\widehat{V}_{off}= \sum'_{i \neq j} V_{ij}'S_{i}^{+}S_{j}^{-}\theta(|\phi_{i}-\phi_{j}|-\phi_{*}) 
\label{eq:Voff}
\end{eqnarray}
 as an off-resonant perturbation that cannot lead to real transitions. Here $\theta(x)=1$ if $x\geq 0$ or $0$ otherwise. 
 This part of interaction will be used below to generate the effective Ising interaction between spins.  Since we choose $\phi_{*}\ll W$ the vast majority of interactions belong to the off-resonant perturbation Eq. (\ref{eq:Voff}).  The remaining resonant interaction 
  \begin{eqnarray} 
\widehat{V}_{res}= \sum'_{i \neq j} V_{ij}'S_{i}^{+}S_{j}^{-}\theta(-|\phi_{i}-\phi_{j}|+\phi_{*})
\label{eq:Vres}
\end{eqnarray}
 should be treated as it is.

In the next step we perform the approximate unitary transformation of the Hamiltonian targeted to remove off-resonant interactions, which is quite similar to the one used in Ref. \cite{Shoekopf,Shhoekopf1}. This transformation suggests the following procedure. The new effective Hamiltonian is created using the transformation
\begin{eqnarray}
H_{eff}=\exp\left(\widehat{S}\right)H\exp\left(-\widehat{S}\right), 
\nonumber\\
\widehat{S}= \sum'_{i \neq j} \frac{V_{ij}S_{i}^{+}S_{j}^{-}}{\phi_{i}-\phi_{j}}\theta(|\phi_{i}-\phi_{j}|-\phi_{*}).  
\label{eq:EffHam}
\end{eqnarray}
The matrix in exponent $\widehat{S}$ is chosen to satisfy the condition \cite{Shoekopf}
\begin{eqnarray}
[\widehat{S}, \widehat{H}_{0}]= -\widehat{V}_{off}.  
\label{eq:Transform}
\end{eqnarray} 
In the case under consideration of a strongly disordered system one can treat $\widehat{S}$ as a perturbation and restrict its consideration to the few lowest order contributions which can be written as (cf. Ref. \cite{Shoekopf})
\begin{eqnarray}
H_{eff}=\widehat{H}_{0}+\widehat{V}_{res}+\widehat{V}^{(2)}+\widehat{V}^{(3)}; 
\nonumber\\
\widehat{V}^{(2)}=\frac{1}{2}\left[\widehat{S}, \widehat{V}_{off}\right]+\left[\widehat{S}, \widehat{V}_{r}\right]; 
\nonumber\\ 
\widehat{V}^{(3)}=\frac{1}{2}\left[\widehat{S}\left[\widehat{S}, \widehat{V}_{r}\right]\right]+\frac{1}{3}\left[\widehat{S}\left[\widehat{S}, \widehat{V}_{off}\right]\right]. 
\label{eq:EffHam1}
\end{eqnarray}
We need the third order contribution because only in this order the effective Ising interaction of spins is generated. The analysis of induced interactions is performed below in Sec. \ref{sec:Corr}.

\subsection{Calculation of induced interactions}
\label{sec:Corr}

\subsubsection{Second order contributions}

Using Eq. (\ref{eq:EffHam1}) one can expressed the interaction induced in the second order in the off-resonant hopping as 
\begin{widetext}
\begin{eqnarray}
\widehat{V}^{(2)}=\widehat{V}_{I}^{(2)}+\widehat{V}_{r}^{(2)}+\widehat{V}_{off}^{(2)};
\nonumber\\
\widehat{V}_{I}^{(2)}=\frac{1}{2}\sum_{i, j}\left[\frac{V_{ij}S_{i}^{+}S_{j}^{-}}{\phi_{i}-\phi_{j}}, V_{ij}S_{j}^{+}S_{i}^{-}\right]=\sum_{i}S_{i}^{z}\sum_{j}\frac{V_{ij}^2}{\phi_{i}-\phi_{j}}; 
\nonumber\\
\widehat{V}_{r}^{(2)}=-\sum_{i \neq j}S_{i}^{+}S_{j}^{-}\theta(\phi_{*}-|\phi_{i}-\phi_{j}|)\sum_{k}\frac{V_{ij}V_{kj}(\phi_{i}+\phi_{j}-2\phi_{k})}{(\phi_{i}-\phi_{k})(\phi_{j}-\phi_{k})}\theta(|\phi_{i}-\phi_{k}|-\phi_{*})\theta(|\phi_{j}-\phi_{k}|-\phi_{*})-
\nonumber\\
\nonumber\\
-2\sum_{i \neq j}S_{i}^{+}S_{j}^{-}\theta(\phi_{*}-|\phi_{i}-\phi_{j}|)\sum_{k}\frac{V_{ij}V_{kj}(\phi_{i}+\phi_{j}-2\phi_{k})}{(\phi_{i}-\phi_{k})(\phi_{j}-\phi_{k})}\theta(\phi_{*}-|\phi_{i}-\phi_{k}|)\theta(|\phi_{j}-\phi_{k}|-\phi_{*}); 
\nonumber\\
\widehat{V}_{off}^{(2)}=-\sum_{i \neq j}S_{i}^{+}S_{j}^{-}\theta(|\phi_{i}-\phi_{j}|-\phi_{*})
\sum_{k}\frac{V_{ij}V_{kj}(\phi_{i}+\phi_{j}-2\phi_{k})}{(\phi_{i}-\phi_{k})(\phi_{j}-\phi_{k})}\theta(|\phi_{i}-\phi_{k}|-\phi_{*})\theta(|\phi_{j}-\phi_{k}|-\phi_{*})-
\nonumber\\
-2\sum_{i \neq j}S_{i}^{+}S_{j}^{-}\theta(|\phi_{i}-\phi_{j}|-\phi_{*})\sum_{k}\frac{V_{ij}V_{kj}(\phi_{i}+\phi_{j}-2\phi_{k})}{(\phi_{i}-\phi_{k})(\phi_{j}-\phi_{k})}\theta(\phi_{*}-|\phi_{i}-\phi_{k}|)\theta(|\phi_{j}-\phi_{k}|-\phi_{*}).
\label{eq:EffHam2}
\end{eqnarray}
\end{widetext}
The results include the corrections to the random field $\widehat{V}_{I}^{(2)}$ and the second order corrections to resonant and off resonant interactions  ($\widehat{V}_{r}^{(2)}$ and $\widehat{V}_{off}^{(2)}$, respectively). The correction to the random field is much smaller than that random field itself. The off-resonant part of interaction can be removed applying one more unitary transformation similarly to Eq. (\ref{eq:Transform}) and it can be used to generate the Ising interaction only in the fourth order in  the hopping. The resonant interaction represents a weak correction to the original resonant interaction Eq. (\ref{eq:Vres}) due to the large strength of disordering compared to the hopping. Thus all these interactions can be approximately neglected. 

\subsubsection{Induced Ising interaction}

\begin{figure}[h!]
\centering
\includegraphics[width=9cm]{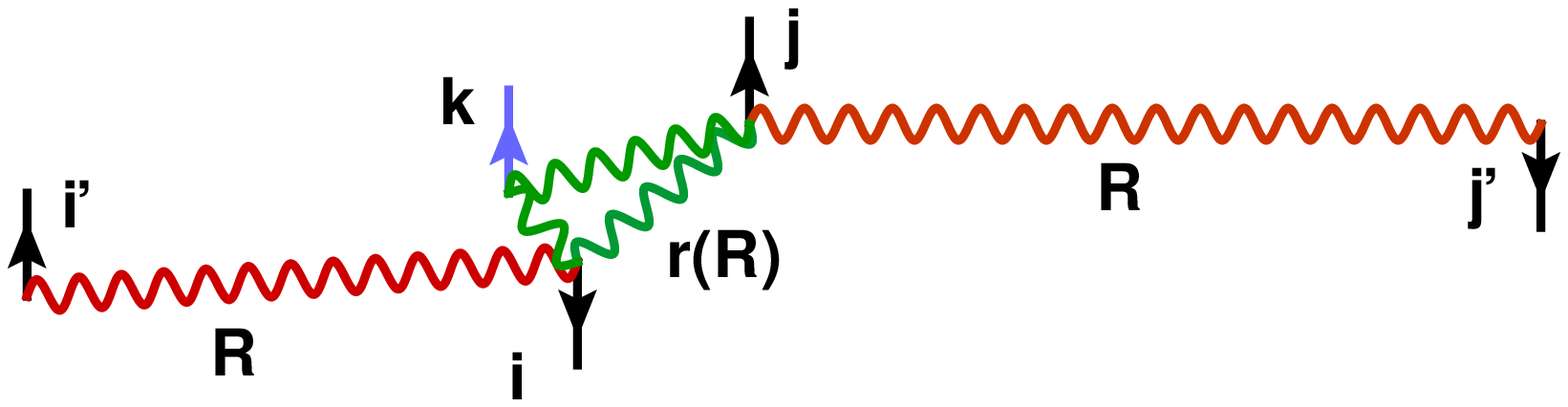}
\caption{Many body interaction of two resonant pairs of spins $(i, i')$ and $(j, j')$ assisted by neighboring spin $k$. Wave-lines indicate significant spin-spin hopping interactions $V$.}
\label{fig:ResPairs}
\end{figure}

The first non-zero contribution to the Ising interaction appears  in the third order in the hopping interaction from three spin loops (see spins $i$, $j$ and $k$ in Fig. \ref{fig:ResPairs}, cf. Ref. \cite{Galperin}). 
We consider only Ising off-resonant contribution to the interaction as the significant third order correction (resonant contribution is smaller as discussed below and off-resonant non-Ising contribution can be removed by the additional unitary transformation). 
This contribution is generated by the last term in the definition of the third order interaction $\widehat{V}^{(3)}$ Eq. (\ref{eq:EffHam1}) and it can be expressed as 
\begin{widetext}
\begin{eqnarray}
\widehat{V}_{I}^{(3)}=\frac{1}{3}\sum_{i, j, k}'\left[\frac{V_{ik}S_{k}^{+}S_{i}^{-}}{\phi_{k}-\phi_{i}}, \left[\frac{V_{jk}S_{j}^{+}S_{k}^{-}}{\phi_{j}-\phi_{k}}, V_{ij}S_{i}^{+}S_{j}^{-}\right]\right]
+\frac{1}{3}\sum_{i, j, k}'\left[\frac{V_{jk}S_{j}^{+}S_{k}^{-}}{\phi_{j}-\phi_{k}}, \left[\frac{V_{ki}S_{k}^{+}S_{i}^{-}}{\phi_{k}-\phi_{i}}, V_{ij}S_{i}^{+}S_{j}^{-}\right]\right],
\label{eq:EffHamV3}
\end{eqnarray}
\end{widetext} 
where the $\sum'$ notation means that the summation is over only off resonant pairs of spins. One can evaluate the commutators in Eq. (\ref{eq:EffHamV3}) as 
\begin{eqnarray}
\left[S_{k}^{+}S_{i}^{-}, \left[S_{j}^{+}S_{k}^{-}, S_{i}^{+}S_{j}^{-}\right]\right]=2S_{j}^{z}(S_{k}^{z}-S_{i}^{z}).
\label{eq:Comm3}
\end{eqnarray}
Then after the straightforward calculations we come up with the following result for the generated Ising interaction of off resonant spins $1/2$ within the $x-y$ model
\begin{eqnarray}
\widehat{U}^{(3)}=\frac{1}{2}\sum_{i,j}'U_{ij}^{(3)}S_{i}^{z}S_{j}^{z}, 
\nonumber\\
U_{ij}^{(3)}=4\sum_{k}'\frac{V_{ij}V_{ik}V_{jk}}{(\phi_{i}-\phi_{k})(\phi_{j}-\phi_{k})}.
\label{eq:ThrdOrdAns}
\end{eqnarray}
There is no such correction in a one-dimensional $XY$ model with the nearest neighbor interactions because it requires spin pairs $i$ and $j$, $j$ and $k$ and $i$, $k$ to be nearest neighbors simultaneously which is not possible in a one-dimension. In other words the product $V_{ij}V_{ik}V_{jk}$ is always equal zero in that model. Therefore it was necessary to introduce the nearest neighbor Ising interaction in Ref. \cite{Pel} to obtain a many-body behavior.

The result Eq. (\ref{eq:ThrdOrdAns}) is valid only if energy differences in denominators are not very small $|\phi_{i}-\phi_{k}|>\phi_{*}$. However the opposite situations occur rarely because of the large random potential strength $W$ compared to the spin-spin interactions.   Indeed, the dominating contribution to the Ising spin-spin interaction of spins $i$ and $j$ comes from the ``assisting'' spins $k$ located in the direct vicinity of either spin $i$ or spin $j$. This is because the interaction is weak ($V_{ik}, V_{jk} \ll W$) and decreases with the distance faster or like $R^{-d}$ (see Ref. \cite{Raikh}), so the sign variable sums over spins $k$ in Eq. (\ref{eq:ThrdOrdAns}) possess the Levy statistics. This means that they are determined by shortest possible distance of order of the average distance between spins.

Typical random field differences for few involved spins are generally given by the energy disorder range $\phi_{i}-\phi_{j} \sim  \phi_{i}-\phi_{k} \sim \phi_{j}-\phi_{k} \sim W$.  Consequently, the Ising spin-spin interaction $U^{(3)}_{ij}S_{i}^{z}S_{j}^{z}$ Eq. (\ref{eq:ThrdOrdAns}) can be estimated as 
\begin{eqnarray}
U_{ij}^{(3)} \approx \frac{V_{ij}^2 \tilde{V}}{W^2} 
\label{eq:IsingInt}
\end{eqnarray}
(remember that  $\tilde{V}$ stands for the nearest neighbor interaction Eq. (\ref{eq:IntHop})). Consequently the induced Ising interaction at the average distance can be estimated as 
\begin{eqnarray}
\tilde{U}^{(3)} \approx \frac{\tilde{V}^3}{W^2}. 
\label{eq:IsAv}
\end{eqnarray}
This result will be used to investigate many-body localization in Sec. \ref{sec:MBLXY} occurring due to the combination of resonant hopping and off-resonant Ising interactions. Below in Sec. \ref{sec:IsingOth} we briefly discuss the induced interaction in other models of interest. 

\subsection{Other models}
\label{sec:IsingOth}

Consider the induced interaction for the problem of Fermi particles Eq. (\ref{eq:HF}). In the case of only non-zero hopping ($U_{ij}=0$) there is no induced interactions  containing the products of two or more operators $n_{i}=a_{i}^{+}a_{i}$. For instance the third order correction to the energy analogous to Eq. (\ref{eq:EffHamV3}) does not induce interaction because the commutators similar to those in Eq. (\ref{eq:Comm3}) lead only to the corrections to a random potential. Indeed one has 
\begin{eqnarray}
\left[a_{k}^{+}a_{i}, \left[a_{j}^{+}a_{k}, a_{i}^{+}a_{j}\right]\right]=n_{i}-n_{j},
\label{eq:Comm4}
\end{eqnarray}
and the corresponding correction to random potentials can be expressed as 
\begin{eqnarray}
\delta \phi^{(3)}_{i}=\sum_{j, k}'\frac{V_{ij}V_{ik}V_{jk}}{(\phi_{i}-\phi_{k})(\phi_{j}-\phi_{k})}. 
\label{eq:FermAns}
\end{eqnarray}
 This is the consequence of the single particle nature of the Fermion problem Eq. (\ref{eq:HF}) with $U_{ij}=0$.  

For the $XY$ model with spin greater than $1/2$ the spin-spin Ising interaction is induced already in the second order of perturbation theory Eq. (\ref{eq:EffHam2}).  This interaction takes the form
\begin{eqnarray}
\widehat{U}=-\frac{1}{2}\sum_{i, j}'\frac{2V_{ij}^{2}S_{i}^{z}S_{j}^{z}(S_{i}^{z}-S_{j}^{z})}{(\phi_{i}-\phi_{j})}. 
\label{eq:HighSpinAns}
\end{eqnarray}
Consequently it depends on the distance as $R_{ij}^{-2\alpha}$ as for the problem with spin $1/2$ Eq. (\ref{eq:IsingInt}). However it is less sensitive to disorder.

\section{Effect of interactions on many-body localization in $XY$ model}
\label{sec:MBLXY}

\subsection{Delocalization in $XY$ model due to long-range interaction}

Consider the joint effect of induced Ising $U_{ij}$ and original hopping $V_{ij}$ interactions on the many-body localization using the previously developed theory  \cite{abmbl1,Lukin2,abmbl2} in the case of the hopping  
and Ising interactions decreasing with the distance according to the power law characterized by exponents $\alpha$ and $\beta=2\alpha$, respectively and by interaction constants given by Eqs. (\ref{eq:IntHop}) and (\ref{eq:IsAv}). 

Several delocalization scenarios have  been considered in Ref. \cite{Lukin2} and the critical dimensions (or critical power law interaction exponents $\alpha$ and $\beta$) for many-body localization have been suggested (we consider only anisotropic interaction case). First two scenarios including hopping  ($\alpha>d$) and small pairs ($\beta>d$) constraints describe the stability of the localized state with respect to the single particle delocalization \cite{PW,Levitov2}. 

The third scenario of ``extended pairs" (see Fig. \ref{fig:ResPairs}) is related to large resonant pairs formed by $XY$ interactions. For each resonant pair of spins $i, i'$ ($S_{i}^{z}+S_{i'}^{z}=0$) the resonant condition \cite{abComment1} 
\begin{eqnarray}
|\phi_{i'}-\phi_{i}|<|V_{ii'}|
\label{eq:ResP}
\end{eqnarray}
is satisfied (in general the perturbation theory corrections to random fields $\phi$ like Eq. (\ref{eq:EffHam2}) comparable to the interaction $V_{ij}$ should be included into field definitions as discussed in Refs. \cite{Levitov2,abmbl1}). Then  two states $S_{i}^{z}=1/2$, $S_{i'}^{z}=-1/2$ and $S_{i}^{z}=-1/2$, $S_{i'}^{z}=1/2$ are represented nearly equally in the system eigenstates. 

The Ising interaction of resonant pairs ($i, i'$) and ($j, j'$) can lead to the energy hopping between them at small energy differences $(\phi_{i}-\phi_{i'})-(\phi_{j}-\phi_{j'})$ compared to their Ising interaction, while the field difference between two pairs $\phi_{i}-\phi_{j}$ can be as large as maximum disorder energy $W$ (in this case the estimate Eq. (\ref{eq:IsingInt}) is well justified). This additional channel of the energy transport leads to the new critical dimension constraint \cite{Lukin2} $d < \frac{\alpha \beta}{\alpha + \beta}$. In the case of $XY$ model one can use the induced interaction Eq. (\ref{eq:IsingInt}) characterized by the exponent $\beta=2\alpha$. Substituting $\beta=2\alpha$ into the extended pair constraint $d < \frac{\alpha \beta}{\alpha + \beta}$ we obtain the new critical dimension constraint for the many-body localization in $XY$ model (see also Table \ref{tbl:scaling})
\begin{eqnarray}
\frac{3d}{2}<\alpha.
\label{eq:CritConstrXY}
\end{eqnarray}

This is the new constraint compared to the previous studies \cite{abmbl1,Lukin2,abmbl2} restricted to the regime of a significant Ising interaction ($\beta\leq \alpha$), where the fourth scenario of  ``Iterated pairs" determines the critical dimension constraint $2d<\beta$ \cite{abmbl1} (anisotropic interaction is assumed). In the case of dominating hopping interaction $\alpha < \beta$ the extended pair scenario gives the stronger restriction of the critical dimension. 

Below we extend the previous consideration to the general problem Eq. (\ref{eq:Hpr}) in the case $\alpha<\beta$ and obtain the  dimension constraint for many-body localization (Sec. \ref{sec:genalphlessbeta}) for arbitrarily relationship of power law interaction exponents $\alpha$ and $\beta$. We also estimate the minimum system size where the localization is still possible at dimension exceeding the critical one (Sec. \ref{sec:maxsize}). The latter study is important for real systems having finite size where the proposed theory can serve as a background for the analysis of the future experimental data. These systems can be realized using the  chains of cold atoms \cite{Lukin1,Lukin2}). The results are summarized in Tables \ref{tbl:scaling} and \ref{tbl:scalingSize}.

\subsection{Dimension constraint for dominating hopping interaction ($\alpha<\beta$)}
\label{sec:genalphlessbeta}

Here we discuss the dimension constraint for the general problem Eq. (\ref{eq:Hpr}) with $\alpha<\beta$ not addressed previously. The induced Ising interaction Eq. (\ref{eq:IsingInt}) can become more significant than the initial Ising interaction if  it decreases with the distance slower than the original Ising interaction i. e. at $2\alpha < \beta$. Therefore considering the dimension constraint one should characterize the Ising interaction by the minimum power law exponent 
\begin{eqnarray}
\beta_{*} = {\rm min} (\beta, 2\alpha), 
\label{eq:betst}
\end{eqnarray}
and use this exponent in the dimension constraint for extended pairs  $d<\frac{\alpha\beta_{*}}{\alpha+\beta_{*}}$ \cite{Lukin2}  as it is given in Table \ref{tbl:scaling}. This result does not depend on the spin since the induced Ising interaction Eq. (\ref{eq:HighSpinAns}) for large spins shows the same distance dependence as for the spin $1/2$ Eq. (\ref{eq:IsingInt}). 

\begin{table}
 \caption{Dimension constraints for  many-body localization for $N \rightarrow \infty$} 
 \label{tbl:scaling}
\centering
\begin{tabular}{|c|c|c|c|}
  \hline
 Model & $\alpha\geq \beta$ & $\alpha<\beta<2\alpha$  & $2\alpha<\beta$ \\
   \hline
  $d_{c}$ & $\frac{\beta}{2}$ & $\frac{\alpha\beta}{\alpha+\beta}$ & $\frac{2\alpha}{3}$ \\ 
   \hline 
   $d_{cF}$ & $\frac{\beta}{2}$ & $\frac{\alpha\beta}{\alpha+\beta}$ & $\frac{\alpha\beta}{\alpha+\beta}$ \\ 
   \hline 
  \end{tabular}
\end{table}

For the Fermi-particles counterpart model with mixed interactions ($\alpha < \beta$) Eq. (\ref{eq:HF}) there is no induced many-body interaction so the extended pair criterion should be always applicable (see Table \ref{tbl:scaling}). 

In the case of violated dimension constraint the delocalization should take place at arbitrarily large strength of disorder beginning from the sufficiently large system size.  We estimate the maximum number of spins where many-body localization can still take place in Sec. \ref{sec:maxsize} below. Although this derivation essentially repeats the arguments of previous work \cite{abmbl1,Lukin2,abmbl2}  there is some qualitative difference of the case $\alpha<\beta$ compared to the previously considered case $\alpha=\beta$ so it can be useful for better understanding of many-body localization breakdown in this regime and for the analysis of the future experimental data in cold atomic systems.  



\begin{widetext}
\begin{table*}
 \caption{Critical strength of disorder and number of particles determining localization threshold for violated dimension constraints Table \ref{tbl:scaling}}.  
 \label{tbl:scalingSize}
\begin{tabular}{|c|c|c|c|}
  \hline
 Model & Specific case & $N_{c}=nR_{*}^{d}$ & $W_{c}$ \\
   \hline
$d \leq \beta < \alpha, 2d$  & $\alpha < \frac{d\beta}{\beta-d}$ & min$\left[\left(\frac{\tilde{V}}{\tilde{U}}\right)^{\frac{d^2}{d(\alpha+\beta)-\alpha\beta}}\left(\frac{W}{\tilde{V}}\right)^{\frac{d\beta}{d(\alpha+\beta)-\alpha\beta}}, \left(\frac{\tilde{V}}{\tilde{U}}\right)^{\frac{d}{2d-\beta}}\left(\frac{W}{\tilde{V}}\right)^{\frac{2d}{2d-\beta}}\right]$ & $\tilde{V}\cdot$max$\left[\left(\frac{\tilde{U}}{\tilde{V}}\right)^{\frac{d}{\beta}} N^{\frac{d(\alpha+\beta)-\alpha\beta}{d\beta}}, \left(\frac{\tilde{U}}{\tilde{V}}\right)^{\frac{1}{2}}N^{\frac{2d-\beta}{2d}}\right]$ \\ 
   \hline    
      & $\frac{d\beta}{\beta-d}<\alpha$ & $\left(\frac{\tilde{V}}{\tilde{U}}\right)^{\frac{d}{2d-\beta}}\left(\frac{W}{\tilde{V}}\right)^{\frac{2d}{2d-\beta}}$ & $\tilde{V}\left(\frac{\tilde{U}}{\tilde{V}}\right)^{\frac{1}{2}}N^{\frac{2d-\beta}{2d}}$ \\ 
   \hline   
$d\leq \alpha=\beta$      & $\beta<2d$  & $\left(\frac{\tilde{V}}{\tilde{U}}\right)^{\frac{d^2}{\beta(2d-\beta)}}\left(\frac{W}{\tilde{V}}\right)^{\frac{d}{2d-\beta}}$ & $\tilde{V}\left(\frac{\tilde{U}}{\tilde{V}}\right)^{\frac{d}{\beta}} N^{\frac{2d-\beta}{d}}$ \\ 
   \hline  
$d\leq \alpha<\beta, \frac{3d}{2}$ & $\beta < \frac{d\alpha}{\alpha-d}$ & min $\left[\left(\frac{\tilde{V}}{\tilde{U}}\right)^{\frac{d^2}{d(\alpha+\beta)-\alpha\beta}}\left(\frac{W}{\tilde{V}}\right)^{\frac{d\beta}{d(\alpha+\beta)-\alpha\beta}}, \left(\frac{W}{\tilde{V}}\right)^{\frac{2d(d+\alpha)}{\alpha (3d -2\alpha)}}\right]$ & $\tilde{V}\cdot$max$\left[\left(\frac{\tilde{U}}{\tilde{V}}\right)^{\frac{d}{\beta}} N^{\frac{d(\alpha+\beta)-\alpha\beta}{d\beta}}, N^{\frac{\alpha (3d -2\alpha)}{2d(d+\alpha)}}\right]$ \\ 
   \hline 
    & $\frac{d\alpha}{\alpha-d}<\beta $ & $\left(\frac{W}{\tilde{V}}\right)^{\frac{2d(d+\alpha)}{\alpha (3d -2\alpha)}}$ & $\tilde{V}N^{\frac{\alpha (3d -2\alpha)}{2d(d+\alpha)}}$\\ 
   \hline     
$\frac{3d}{2}<\alpha<\beta$ & $\beta < \frac{d\alpha}{\alpha-d}$ & $\left(\frac{\tilde{V}}{\tilde{U}}\right)^{\frac{d^2}{d(\alpha+\beta)-\alpha\beta}}\left(\frac{W}{\tilde{V}}\right)^{\frac{d\beta}{d(\alpha+\beta)-\alpha\beta}}$ &  $\tilde{V}\left(\frac{\tilde{U}}{\tilde{V}}\right)^{\frac{d}{\beta}} N^{\frac{d(\alpha+\beta)-\alpha\beta}{d\beta}}$ \\ 
   \hline      
  \end{tabular}
\end{table*}
\end{widetext}

\subsection{Critical size and disorder strength at large dimension $d>d_{c}$}
\label{sec:maxsize}

We begin with the analysis of delocalization in the general case $\alpha<\beta$ including $XY$ model. Assume that the dimension constraint for extended pairs $d<\frac{\alpha\beta_{*}}{\alpha+\beta_{*}}$ is not satisfied so the delocalization should take place at arbitrarily large strength of disorder at sufficiently large system size. Our targets are the dependence of critical disorder $W_{c}$ on the system size $L$ expressed through the number of spins $N=nL^{d}$ and the dependence of the number of spins $N_{c}$ needed for the delocalization to occur at the given strength of  disorder $W$. 
Our consideration follows the previous analysis of similar problems for $\alpha \geq \beta$ \cite{abmbl1,abmbl2}.

The dynamics of strongly disordered system is primarily associated with resonant pairs Eq. (\ref{eq:ResP}) where the random field energy can be compensated by the hopping interaction. The resonant pairs can be divided into sub classes of typical size $R$ and typical energy $V(R)=\frac{\tilde{V}}{(n^{\frac{1}{d}}R)^{\alpha}}$. The density of resonant pairs having typical size $R$ is given by \cite{abmbl2}
\begin{eqnarray}
n(R) = n \frac{\tilde{V}}{W}\frac{1}{(n^{\frac{1}{d}}R)^{\alpha-d}}. 
\label{eq:RPDens1}
\end{eqnarray}  

Ising interaction characterized by the interaction strength at the average distance $\tilde{U}$ creates the dynamic interaction of resonant pairs of the typical size $R$ leading to the simultaneous flip flop transition in both resonant pairs accompanied by the energy hop between them. Since both pairs are resonant the amplitude of this simultaneous transition is of order of their bare Ising interaction which can be estimated as \cite{abmbl2}
\begin{eqnarray}
U(R) = U n(R)^{\frac{\beta}{d}} \tilde{U}  \left(\frac{\tilde{V}}{W}\right)^{\frac{\beta}{d}}\frac{1}{(n^{\frac{1}{d}}R)^{\frac{\beta(\alpha-d)}{d}}}. 
\label{eq:RPInt}
\end{eqnarray} 
One should notice that in the case of extended pairs and $\alpha < 2d$ the average distance between pairs $n(R)^{-\frac{1}{d}}$ becomes smaller than their size at sufficiently large size $R$ (see Ref. \cite{abmbl2}, Eq. (\ref{eq:RPDens1})) for two resonant pairs. Therefore the coupling of only the closest spins is important as shown in Fig. \ref{fig:ResPairs}. Then the isotropic or anisotropic character of Ising interaction does not affect the estimate of Ising interaction between pairs Eq. (\ref{eq:RPInt}) and consequently the delocalization transition in contrast with the case $\alpha > \beta$ of dominating Ising interaction considered in Ref. \cite{Lukin2}.  

The interaction of resonant pairs Eq. (\ref{eq:RPInt}) represents their dynamic coupling that can lead to delocalization if it exceeds the typical energy $V(R)$ of the subsystem of pairs. Considering power law distance dependencies of $U(R)$ and $V(R)$ one can see that  if the criterion $d \leq \frac{\alpha\beta}{\alpha+\beta}$ is not satisfied then the interaction of pairs of the size $R$ ($U(R)$) decreases with the distance slower than the typical energy of those pairs ($V(R)$). Therefor inevitable delocalization is expected at sufficiently large system size where the pair coupling exceeds their typical energy. The size $R_{*}$ where two energies are equal to each other estimates the maximum  system size where many-body localization is still possible  at the given strength of disorder $W$. Solving the equation $U(R_{*}) = V(R_{*})$ we estimate this size as  
\begin{eqnarray}
R_{*}=n^{-\frac{1}{d}}\left(\frac{\tilde{V}}{\tilde{U}}\right)^{\frac{d}{d(\alpha+\beta)-\alpha\beta}}\left(\frac{W}{\tilde{V}}\right)^{\frac{\beta}{d(\alpha+\beta)-\alpha\beta}}. 
\label{eq:MinSIze}
\end{eqnarray}
In the case of $XY$ model with the spin $1/2$ the Ising interaction strength $\tilde{U}$ depends itself on the strength of disorder Eq. (\ref{eq:IsingInt}). In this case the dependence of the maximum system size on disorder Eq. (\ref{eq:MinSIze}) should be modified as (remember that $\beta=2\alpha$ in that case)
\begin{eqnarray}
R_{*XY}=n^{-\frac{1}{d}}\left(\frac{W}{\tilde{V}}\right)^{\frac{2(d+\alpha)}{\alpha (3d -2\alpha)}}. 
\label{eq:MinSIzeXY}
\end{eqnarray}

Consider the general model with mixed interaction Eq. (\ref{eq:Hpr}) for $\alpha<\beta$.  If both constraints for mixed interactions ($\alpha\beta/(\alpha+\beta)>d$) and for $XY$ model ($\alpha<3d/2$) are violated then one  should choose the strongest interaction out of two leading to the lowest size constraint determined by the minimum of two estimates in  Eqs. (\ref{eq:MinSIze}) and (\ref{eq:MinSIzeXY}). If only one of two conditions is satisfied then one should use the corresponding criterion. The estimates of the critical number of spins $N=nR_{*}^{d}$ in all these cases is given in Table \ref{tbl:scalingSize}. 

Case $\alpha = \beta$ considered in Ref. \cite{abmbl2} can be described by Eq. (\ref{eq:MinSIze}) setting $\alpha=\beta$. In the remaining case of $\alpha<\beta$ the iterated pair scenario has been suggested \cite{Lukin2} (see also \cite{abmbl1}). In that scenario the delocalization is determined by interacting nearest neighbor resonant pairs which replace spins in the original model. For anisotropic interaction \cite{Lukin2} the problem can be reformulated with the modified parameters $W_{p}=\tilde{V}$, $n_{p}=n\tilde{V}/W$, $\tilde{U}_{p}=\tilde{V}_{p}=\tilde{U}(\tilde{V}/W)^{\frac{\beta}{d}}$ and $\alpha_{p}=\beta_{p}=\beta$. Using these parameters in Eq. (\ref{eq:MinSIze}) one can estimate the critical size assuming $\beta<2d$ as 
\begin{eqnarray}
R_{*P}=n^{-\frac{1}{d}}\left(\frac{W}{\tilde{V}}\right)^{\frac{2}{2d-\beta}}\left(\frac{\tilde{V}}{\tilde{U}}\right)^{\frac{1}{2d-\beta}}. 
\label{eq:MinSIzealphLarge}
\end{eqnarray} 
If both extended pair and iterated pair criteria lead to the delocalization ($\alpha>\beta$, $\beta<2d$, $\alpha\beta/(\alpha+\beta)<d$) then one should choose the  minimum size determined by either  Eq. (\ref{eq:MinSIze}) or Eq. (\ref{eq:MinSIzealphLarge}) corresponding to the most efficient delocalization.

In Table \ref{tbl:scaling} we show the estimates of the maximum number of spins ($N_{*}=nR_{*}^{d}$) where many-body localization is still possible for all possible exponents $\alpha$ and $\beta$ and violated dimension constraints. The critical strength of disorder $W_{c}$ for the given number of spins can be found resolving Eqs. (\ref{eq:MinSIze}), (\ref{eq:MinSIzeXY}), (\ref{eq:MinSIzealphLarge}) with respect to the disorder $W$ at fixed system size $R$ or number of spins $N$. The dependencies $W_{c}(N)$ are also shown in Table \ref{tbl:scalingSize}. The results can be easily generalized for the interacting Fermions simply removing the parts related to the induced Ising interaction which does not exist in that case.

\subsection{Threshold regimes $d=d_{c}$}
\label{sec:alph=d}

The threshold regimes where interactions behavior corresponds to the border line of the dimension constraints need the special attention. There are two distinguishable regimes including the cases $\alpha=d$ or $\beta=d$,  where a single particle delocalization becomes important at very large system size for arbitrarily large strength of disorder \cite{Levitov2,Lukin2} and the cases $\beta=2d$, $\alpha=3d/2$ and $\alpha\beta/(\alpha+\beta)=d$ corresponding to the border line for many-body localization breakdown. The first regime is particularly significant because of the common appearance of $R^{-d}$ interactions due to dipolar or elastic forces \cite{Yu,YuLegett}. 

We believe that the single-particle delocalization can be ignored for $\alpha=d$ (and similarly for $\beta=d$) in the regime of strong disorder $\tilde{V} \ll W$ because the delocalization becomes significant only if the logarithmic delocalization parameter $\frac{\tilde{V}}{W}\ln\left(N\right)$ approaches unity \cite{Levitov2}. This can happen only at exponentially large system size remarkably exceeding the critical sizes from Table \ref{tbl:scalingSize} for many-body localization. Therefore we do not expect any related changes in our estimates for critical size and disorder behaviors except for may be additional weak logarithmic parametric dependencies which are beyond the scope of our qualitative analysis. 

The number of resonant interactions grows logarithmically with the system size at the many-body localization thresholds ($\beta=2d$, $\alpha=3d/2$ or $\alpha\beta/(\alpha+\beta)=d$). This behavior is similar to the single particle localization problem with strong disorder and $R^{-d}$ hopping interaction \cite{Levitov2}. In this regime  the excitations are nearly localized: their typical localization radius is much smaller than the system size, but it approaches infinity in the infinite system limit (see however Ref. \cite{Levitov4} where delocalization can take place under certain conditions at exponentially large but finite radius). In our case we similarly expect that the system remains localized until the system size gets exponentially large $\ln(n^{\frac{1}{d}}R) \sim W/\tilde{V}$ while we are not able to describe the system behavior at larger sizes. Perhaps the experimental investigation of the implementations of this model in cold atomic gases as suggested in Refs. \cite{Lukin1,Lukin2} will help to resolve this difficult issue.

Below in Sec. \ref{sec:num} the predictions of theory for the $XY$ model are verified for one-dimensional model using the numerical finite size scaling of the ergodicity parameter \cite{abmbl2}. 

\section{Finite size scaling for $XY$ model}
\label{sec:num}


Consider  a random one dimensional $XY$  model. In the numerical study we describe the localization-delocalization transition using the ergodicity parameter \cite{BGold,abmbl2}. In the thermodynamic limit of infinite system this parameter is defined as the local spin-spin correlation function 
\begin{eqnarray}
Q=\lim_{t\rightarrow \infty}\frac{\sum_{m}4<m|S_{i}^{z}(t)S_{i}^{z}(0)|m>\delta(E_{m})}{\sum_{m}\delta(E_{m})}
\label{eq:BGpar}
\end{eqnarray}
taken in the infinite time limit \cite{BGold} and averaged over the  system eigenstates having zero energy. The latter choices correspond to  the infinite temperature thermodynamic limit \cite{OgHuse1,Reichmann}. In that limit the average ergodicity parameter  should approach zero in the delocalized state where correlations are subject to decay, while  in the localized state it should be finite (unity in an infinitely strong disorder limit). Our method is perhaps not as efficient in the definition of the localization transition point as recently developed analysis of entanglement entropy \cite{OgHuse1,Dresden}; yet it permits the easy determination of the scaling of the localization transition with the system size that can be compared to the theory predictions in Table \ref{tbl:scalingSize}. 

The numerical study  is performed for the Hamiltonian Eq. (\ref{eq:H}) describing the periodic one dimensional chain of $N$ spins separated by the unit distance  using the periodic interaction 
\begin{eqnarray}
V_{ij} = \pm\frac{1}{max(|i-j|, N-|i-j|)^{\alpha}} 
\label{eq:numint}
\end{eqnarray}
with random, uncorrelated signs of all interactions. This model is similar to the models studied in Ref. \cite{Lukin2,abmbl2} with the only difference that the Ising spin-spin interaction is set equal to zero here. The total spin projection to the $z$ axis is conserved. We restrict the numerical consideration to the states with zero total spin projection which represents well the thermodynamic limit. 

In the numerical study the ergodicity parameter Eq. (\ref{eq:BGpar}) is defined as a configuration averaged spin-spin correlation function at infinite time \cite{BGold} over the narrow band of $N_{\alpha}$ eigenstates $\alpha$ of the Hamiltonian Eq. (\ref{eq:H}) with zero total spin and  energy $E_{\alpha}$ belonging to the domain $(-\delta <E_{\alpha}< \delta)$ 
\begin{eqnarray}
Q = \frac{4}{N_{\alpha}}\sum_{\alpha}\mid <\alpha|\delta S^{z}|\alpha>|^2 \theta(\delta-|E_{\alpha}|). 
\label{eq:Q}
\end{eqnarray}
The domain size $\delta =0.04W\sqrt{N}$  is determined requiring the many-body density of states 
\begin{eqnarray}
g(E) \approx \frac{\exp\left(-\frac{E^2}{24NW^2}\right)}{\sqrt{24\pi NW^2}} 
\label{eq:MBDoS}
\end{eqnarray}
to change at the scale $\delta$ by $1\%$.  The function $g(E)$ has been estimated using the random potential part of the system Hamiltonian Eq. (\ref{eq:H0}) 
employing the law of large numbers ($N \gg 1$) and large strength of disordering $\tilde{V}\ll W$. 

 \begin{figure}[h!]
\centering
\includegraphics[width=9cm]{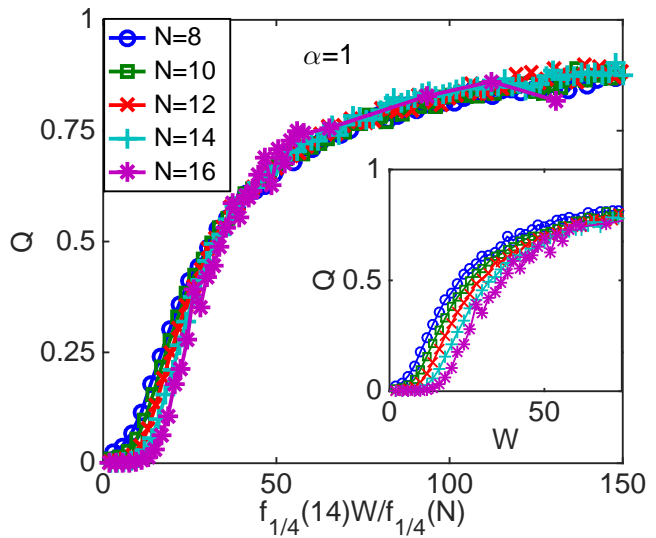}
\caption{Original (inset) and rescaled dependencies of ergodicity parameter $Q$ on  disordering $W$ for $\alpha=1$.}
\label{fig:a1}
\end{figure}

\begin{figure}[h!]
\centering
\includegraphics[width=9cm]{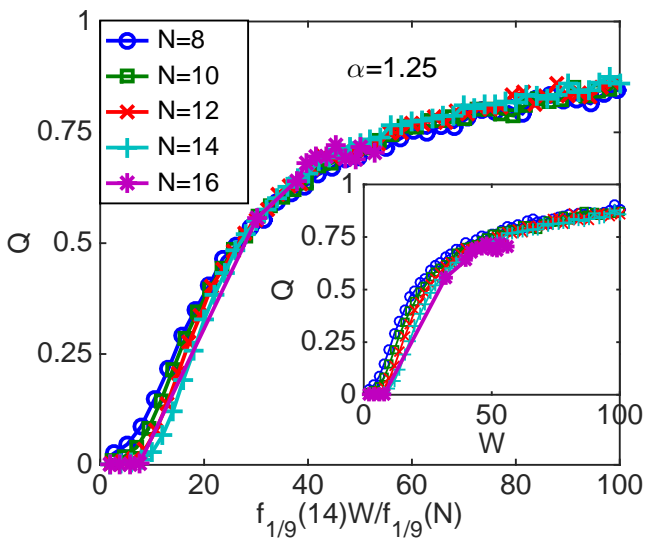}
\caption{Original (inset) and rescaled dependencies of ergodicity parameter $Q$ on  disorder $W$ for  $\alpha=1.25$.}
\label{fig:a1_25}
\end{figure}

\begin{figure}[h!]
\centering
\includegraphics[width=9cm]{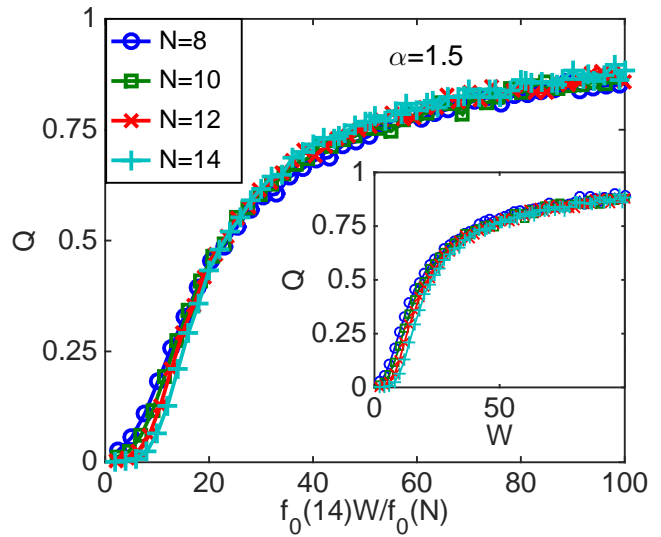}
\caption{Original (inset) and rescaled dependencies of ergodicity parameter $Q$ on  disorder $W$ for  $\alpha=1.5$.}
\label{fig:a1_5}
\end{figure}

\begin{figure}[h!]
\centering
\includegraphics[width=9cm]{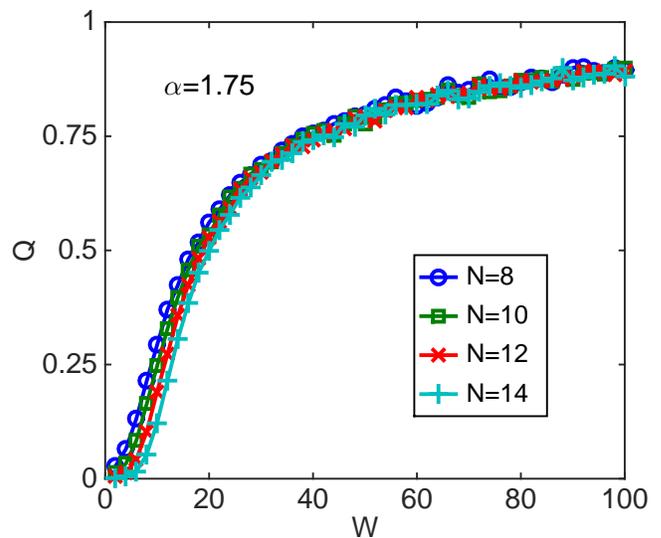}
\caption{Dependence of ergodicity parameter $Q$ on  disorder $W$ for  $\alpha=1.75$.}
\label{fig:a1_75}
\end{figure}

\begin{figure}[h!]
\centering
\includegraphics[width=9cm]{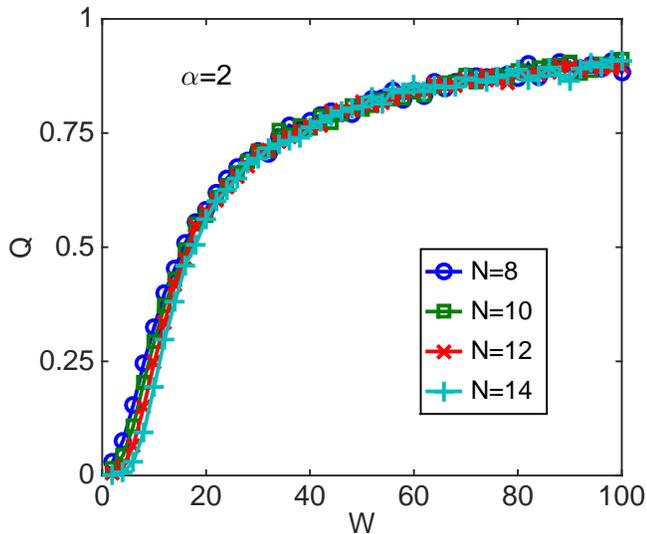}
\caption{Dependence of ergodicity parameter $Q$ on  disorder $W$ for  $\alpha=2$.}
\label{fig:a2}
\end{figure}

The parameter $Q$ has been averaged over few  thousands realizations of the system Hamiltonian to have the relative error of $1\%$ ($5\%$ for $N=16$). The $XY$ models with interactions characterized by power law exponents $\alpha= 1, 1.25, 1.5, 1.75, 2$ taken around the predicted threshold $\alpha=3/2$ (see Table \ref{tbl:scaling}) have been considered for a strength of disorder $W$ ranging between $W=1$ and $W=150$ (we set $\tilde{V}=n=1$). In all systems the localization transition can be seen with increasing disorder; i. e. $Q$ approaches $0$ at small $W$ and tends to $1$ in the opposite limit of $W \rightarrow \infty$ (see Figs. \ref{fig:a1}, \ref{fig:a1_25}, \ref{fig:a1_5}, \ref{fig:a1_75}, \ref{fig:a2}). However these transitions behave differently for different exponents $\alpha$ and numbers of spins $N$.

The results for the dependence of ergodicity parameter on the strength of disorder  are shown in the insets of Figs. \ref{fig:a1}, \ref{fig:a1_25}, \ref{fig:a1_5} for $\alpha \leq 3/2$ and in Figs. \ref{fig:a1_75} and \ref{fig:a2} for $\alpha=1.75$ and $2$, respectively. For $\alpha < 3/2$ the transition shifts towards large disorder $W$ with increasing the number of spins, while for $\alpha > 3/2$ the transition is almost insensitive to the system size. The situation is inconclusive for $\alpha=3/2$ where larger system sizes should be probed. The result of the visual inspection  agrees with the qualitative expectation for the threshold power law exponent $\alpha_{c}=3/2$ separating the regimes where the delocalization is inevitable in the large $N$ limit ($\alpha < 3/2$) and where it is possible.  

For quantitative characterization of the critical strength of disorder $W_{c}$ dependence on the number of spins $N$ we use the scaling function \cite{abmbl2}
\begin{eqnarray}
f_{a}(N)=\frac{N}{2(N-1)}\left(2\sum_{n=1}^{N/2-1}n^{a-1}+\left(\frac{2}{N}\right)^{a-1}\right). 
\label{eq:sc_func}
\end{eqnarray} 
This function replaces the continuous power law dependence $N^{a}$ predicted by the qualitative consideration (Table \ref{tbl:scalingSize}) with the discrete sum of resonance probabilities which accounts better for finite size effects. For $a>0$ this function has the  asymptotic behavior $f_{a}(N) \propto N^{a}$. 

In Figs. \ref{fig:a1}, \ref{fig:a1_25}, \ref{fig:a1_5} we show the modified plot of the data for $\alpha \leq 3/2$ with disorder strength rescaled using the function Eq. (\ref{eq:sc_func}) with the theoretically predicted exponents $a=(\alpha-3d/2)/(1+\alpha)$ (see Eq. (\ref{eq:MinSIzeXY}) and Table \ref{tbl:scalingSize}). This rescaling leads to a reasonable match between the graphs for different numbers of spins $N$ indicating a good agreement of theoretically predicted and numerically found scaling of the transition point with the system size. 

 \begin{figure}[h!]
\centering
\includegraphics[width=9cm]{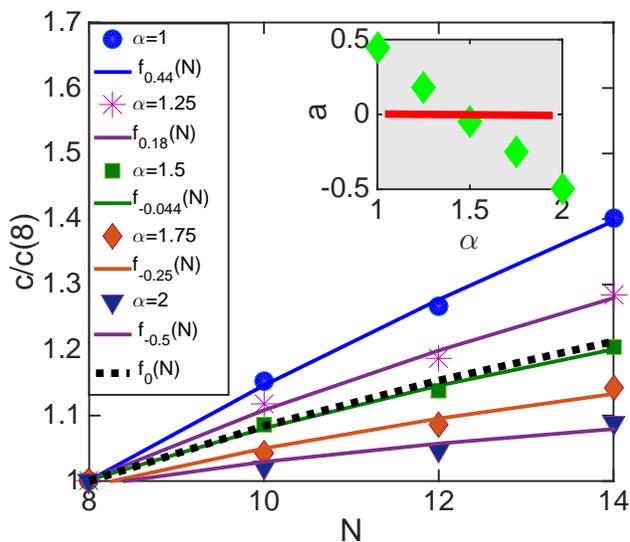}
\caption{Scaling factors $c$ vs. a number of spins $N$ for different interactions ($\alpha$) and the fit of $c(N)$ dependencies using  functions $f_{a}(N)$ with optimum fitting parameters $a$ estimating scaling exponents for localization transition size dependence.  Inset shows the dependence of exponents $a$ on the interaction power law exponents $\alpha$. The solid (red) line indicates the threshold case $a=0$. }
\label{fig:Scal}
\end{figure}

In addition to the visual inspection  we performed  quantitative estimates of rescaling parameters for the data sets $Q_{N}(W)$ with the same exponent $\alpha$ and different $N's$ using the optimization procedure  
\cite{abmbl2}. For each exponent $\alpha$ we determined the set of optimum rescaling parameters $c_{\alpha}(N)$  corresponding to the minimum  of  the squared deviation $\sum_{i}(Q_{14}(W_{i})-Q_{N}(c_{\alpha}(N)W_{i}))^{2}$ with $W_{i}$ changing with the step of $1$ from $1$ to $100$. The results for different rescaling parameters $c(\alpha, N)$ are shown in Fig. \ref{fig:Scal} by symbols defined within the graph. 

To characterize the change of the rescaling factors and consequently the critical strength of disorder $W_{c}$ with the number of spins we fitted each data set by the function $f_{a}(N)$ Eq. (\ref{eq:sc_func}) choosing the parameter $a$ to attain the best agreement of the model with the numerical results. The solid lines in Fig. \ref{fig:Scal} show these optimum fits and the exponents $a$ are shown in the inset and indicated in the legend. The scaling function $f_{0}(N)$ serves as the crossover between the regimes of finite ($a<0$) and infinite ($a>0$) localization thresholds in the thermodynamic limit $N\rightarrow \infty$.  In agreement with the theoretical expectations of the crossover at $\alpha=3/2$ we found $a>0$  for  $\alpha=1, 1.25$, $a<0$ for $\alpha = 1.75, 2$, while situation is not clear at the threshold $\alpha=1.5$ as shown in the inset to Fig. \ref{fig:Scal}.    The estimated exponents $a$ are, however, somewhat larger than the theory predictions Eq. (\ref{eq:MinSIzeXY}) which can be due to finite size effects. The logarithmic increase of the number of resonant interactions with increasing the number of spins  for $\alpha=1$ (see Ref. \cite{Levitov2} and Sec. \ref{sec:alph=d})  can lead to the observed deviations.

\section{Conclusion}
\label{sec:Conclusion}

We investigated  many-body localization in the $XY$ model with the long-range interaction  $V(R)\propto R^{-\alpha}$ in a strongly disordered  regime. In this regime  the Ising interaction decreasing with the distance as $R^{-2\alpha}$ appears in the third order of perturbation theory with respect to the hopping interaction.  Considering the combined effect of this induced interaction and the original hopping interaction we found the novel dimension constraint $\alpha > 3d/2$ required to attain the many body localization in the thermodynamic limit of infinite system. 

Using this result we suggested the dimension constraints for the general problem of many-body localization in the presence of the long-range hopping and Ising interactions with arbitrarily power law distance dependencies (see Table \ref{tbl:scaling}). In the case when the dimension constraint is violated the maximum system size where the localization is still possible is predicted (see Table \ref{tbl:scalingSize}). Also we predicted the critical disorder dependence on the systems size. These results can be used to interpret the recently proposed experiments in cold atomic systems implementing the disordered spin systems with the long-range interactions \cite{Lukin1,Lukin2}. 

It turns out that many-body localization in $XY$ model is more stable with respect to the long-range interaction then in the model of spins having both Ising and $XY$ long-range interactions \cite{abmbl2}.  For instance many-body localization can be attained in a thermodynamic limit in a three dimensional $XY$ model with a quadrupole  interaction ($\alpha =5$) while it is not possible in the spin system with a quadrupole Ising and $XY$ interactions.  Thus  spin systems with only $XY$ interaction can be more attractive for quantum informatics. 

It is interesting to notice that the effective long-range interaction of spins can be generated in periodically driven interacting systems \cite{first,Lazarides,Roy,Ponti} as a part of effective Hamiltonian as suggested in Ref. \cite{Roy}. Therefore the scaling behaviors similar to the ones studied in the present work can be seen in these systems as well (e. g. the scaling of the decay time with drive parameters \cite{Roy}). The detail analysis of the relationship between two problems is beyond the scope of the present work and will be performed later. 

The author acknowledges Louisiana EPSCORE LA Sigma and LINK Programs for support.


\begin{thebibliography}{00}
\bibitem{Comment1}
The low temperature regime $T < W/k_{B}$ in the case of strong disorder for the problem described by the Hamiltonian Eq. (\ref{eq:Hpr}) (higher temperatures can be approximately described by the infinite temperature case) can be studied  restricting the relevant spins to those which are not frozen, i. e. having the effective random field smaller than the thermal energy $|\phi_{i}|< k_{B}T$. This subsystem of thermal spins can be identified removing gradually high energy ``frozen" spins with large random fields $|\phi_{i}| \gg k_{B}T$ and replacing their interactions with other spins with the random fields acting on those remaining spins. The relevance  of this procedure is justified by the weakness of the interaction of spins in a strongly disordered system under consideration. Indeed, it is much smaller than the typical spin energy $\tilde{U}, \tilde{V} \ll W$ at high temperature and the reduction of typical energy of spins by some factor $\eta>1$ leads to the reduction of the characteristic interactions of spins by the factors of $\eta^{\alpha/d}$ or $\eta^{\beta/d}$ making interaction even smaller compared to disorder strength in the case under consideration $\alpha, \beta \geq d$ (in the opposite case the low temperature state needs a special consideration; for instance if $\beta<d$ the pseudogap in the density of states is formed \cite{ESbook}). Then the dynamics of thermal spins having random energy comparable to the thermal energy can be approximately treated using the infinite temperature consideration. For the system of these ``thermal'' spins the model similar to Eq. (\ref{eq:Hpr}) can be extended replacing disorder with the thermal energy ($W \rightarrow k_{B}T$), density of spins with the density of thermal spins ($n \rightarrow nk_{B}T/W$) and interactions at average distances to interactions at average distances between thermal spins ($\tilde{V} \rightarrow \tilde{V} (k_{B}T/W)^{\frac{\alpha}{d}}$, $\tilde{U} \rightarrow \tilde{U} (k_{B}T/W)^{\frac{\beta}{d}}$, see also Refs.  \cite{abmbl1,abmbl2}).
\bibitem{abComment1}
Random fields $\phi_{i}$ and $\phi_{i'}$  are taking with all proper corrections of perturbation theory (e. g. Eq. (\ref{eq:EffHam2}) which converges quite well in the regime of a large strength of disorder \cite{abmbl1}.
\end{thebibliography}
\end{document}